\documentclass[12pt]{article}

\usepackage[body={17.5cm, 21cm},right=2cm]{geometry}

\usepackage{booktabs}
\usepackage{caption}

\usepackage{color}
\usepackage{graphicx}
\usepackage{epsf}
\usepackage{graphicx,epsfig}
\usepackage{multirow}

\usepackage{amsmath, amsthm, amssymb}

\usepackage{epsfig}
\usepackage{cite}
\usepackage{color,colordvi}
\newcommand{\be}{\begin{eqnarray}}
\newcommand{\ee}{\end{eqnarray}}
\newcommand{\bi}{\begin{itemize}}
\newcommand{\ei}{\end{itemize}}

\newcommand{\bx}{{\bf{x}}}

\newcounter{hran}

\makeatletter
\renewcommand\section{\@startsection {section}{1}{\z@}%
                               {-3.5ex \@plus -1ex \@minus -.2ex}%
                               {2.3ex \@plus.2ex}%
                               {\normalfont\large\bfseries}}
\makeatother

\def\MSbar{\relax\ifmmode\overline{\rm MS}\else{$\overline{\rm MS}${ }}\fi}

\def\d{\rm d}

\def\tr{{\rm tr}}

\numberwithin{equation}{section}
\begin{document}
\vspace{5mm}
\vspace{0.5cm}

\begin{center}

\def\thefootnote{\fnsymbol{footnote}}

{\large \bf 
  Non-Linear  Single Higgs MSSM  
  %
}
\\[1.5cm]
{\large  Fotis Farakos and Alex Kehagias }
\\[0.5cm]

\vspace{.3cm}
{\normalsize {\it  Physics Division, National Technical University of Athens, \\15780 Zografou Campus, Athens, Greece}}\\

\vspace{.3cm}
{\normalsize { E-mail: fotisf@mail.ntua.gr,  kehagias@central.ntua.gr }}


\end{center}

\vspace{3cm}

\hrule \vspace{0.3cm}
{\small  \noindent \textbf{Abstract} \\[0.3cm]
\noindent 
We present a non-linear  MSSM with non-standard Higgs sector and goldstino field.
Non-linear supersymmetry for the 
goldstino couplings is described by the constrained chiral superfield and, 
as usual, the Standard Model sector is encompassed in suitable chiral and vector supermultiplets. 
Two models are presented.  In the first model (non-linear MSSM$3\frac{1}{2}$), 
the second Higgs  is replaced by a new supermultiplet of 
half-family with only a new generation of leptons (or quarks).
In the second model, for anomaly cancellation purposes, the second Higgs 
 is retained as a spectator superfield by imposing a discrete symmetry. 
 Both models do not have a $\mu$-problem as 
a $\mu$-term is forbidden by the discrete symmetry in the case of a spectator second Higgs or not 
existing at all in the case of a single Higgs. 
Moreover, the tree level relation between the Higgs mass and the hidden sector SUSY 
breaking scale $\sqrt{f}$ is derived. Finally, we point out a relative suppression 
by $m_{soft}/\Lambda$ of the bottom and tau Yukawa couplings with respect 
to those of the top quark.

\vspace{0.5cm}  \hrule
\vskip 1cm

\def\thefootnote{\arabic{footnote}}
\setcounter{footnote}{0}



\baselineskip= 19pt

\newpage 

\newcommand{\fix}{\Phi(\mathbf{x})}
\newcommand{\fiLx}{\Phi_{\rm L}(\mathbf{x})}
\newcommand{\fiNLx}{\Phi_{\rm NL}(\mathbf{x})}
\newcommand{\fik}{\Phi(\mathbf{k})}
\newcommand{\fiLk}{\Phi_{\rm L}(\mathbf{k})}
\newcommand{\fiLkone}{\Phi_{\rm L}(\mathbf{k_1})}
\newcommand{\fiLktwo}{\Phi_{\rm L}(\mathbf{k_2})}
\newcommand{\fiLkthree}{\Phi_{\rm L}(\mathbf{k_3})}
\newcommand{\fiLkfour}{\Phi_{\rm L}(\mathbf{k_4})}
\newcommand{\fiNLk}{\Phi_{\rm NL}(\mathbf{k})}
\newcommand{\fiNLkone}{\Phi_{\rm NL}(\mathbf{k_1})}
\newcommand{\fiNLktwo}{\Phi_{\rm NL}(\mathbf{k_2})}
\newcommand{\fiNLkthree}{\Phi_{\rm NL}(\mathbf{k_3})}

\newcommand{\kernel}{f_{\rm NL} (\mathbf{k_1},\mathbf{k_2},\mathbf{k_3})}
\newcommand{\dirac}{\delta^{(3)}\,(\mathbf{k_1+k_2-k})}
\newcommand{\dirackonektwokthree}{\delta^{(3)}\,(\mathbf{k_1+k_2+k_3})}

\newcommand{\beq}{\begin{equation}}
\newcommand{\eeq}{\end{equation}}
\newcommand{\beqarr}{\begin{eqnarray}}
\newcommand{\eeqarr}{\end{eqnarray}}

\newcommand{\angk}{\hat{k}}
\newcommand{\angn}{\hat{n}}

\newcommand{\tfnow}{\Delta_\ell(k,\eta_0)}
\newcommand{\tf}{\Delta_\ell(k)}
\newcommand{\tfone}{\Delta_{\el\ell_1}(k_1)}
\newcommand{\tftwo}{\Delta_{\el\ell_2}(k_2)}
\newcommand{\tfthree}{\Delta_{\el\ell_3}(k_3)}
\newcommand{\tffour}{\Delta_{\el\ell_1^\prime}(k)}
\newcommand{\deltatilde}{\widetilde{\Delta}_{\el\ell_3}(k_3)}

\newcommand{\alm}{a_{\ell m}}
\newcommand{\almL}{a_{\ell m}^{\rm L}}
\newcommand{\almNL}{a_{\ell m}^{\rm NL}}
\newcommand{\almone}{a_{\el\ell_1 m_1}}
\newcommand{\almLone}{a_{\el\ell_1 m_1}^{\rm L}}
\newcommand{\almNLone}{a_{\el\ell_1 m_1}^{\rm NL}}
\newcommand{\almtwo}{a_{\el\ell_2 m_2}}
\newcommand{\almLtwo}{a_{\el\ell_2 m_2}^{\rm L}}
\newcommand{\almNLtwo}{a_{\el\ell_2 m_2}^{\rm NL}}
\newcommand{\almthree}{a_{\el\ell_3 m_3}}
\newcommand{\almLthree}{a_{\el\ell_3 m_3}^{\rm L}}
\newcommand{\almNLthree}{a_{\el\ell_3 m_3}^{\rm NL}}

\newcommand{\YLMstar}{Y_{L M}^*}
\newcommand{\Ylmstar}{Y_{\ell m}^*}
\newcommand{\Ylmstarone}{Y_{\el\ell_1 m_1}^*}
\newcommand{\Ylmstartwo}{Y_{\el\ell_2 m_2}^*}
\newcommand{\Ylmstarthree}{Y_{\el\ell_3 m_3}^*}
\newcommand{\Ylmstarfour}{Y_{\el\ell_1^\prime m_1^\prime}^*}
\newcommand{\Ylmstarfive}{Y_{\el\ell_2^\prime m_2^\prime}^*}
\newcommand{\Ylmstarsix}{Y_{\el\ell_3^\prime m_3^\prime}^*}

\newcommand{\YLM}{Y_{L M}}
\newcommand{\Ylm}{Y_{\ell m}}
\newcommand{\Ylmone}{Y_{\el\ell_1 m_1}}
\newcommand{\Ylmtwo}{Y_{\el\ell_2 m_2}}
\newcommand{\Ylmthree}{Y_{\el\ell_3 m_3}}
\newcommand{\Ylmfour}{Y_{\el\ell_1^\prime m_1^\prime}}
\newcommand{\Ylmfive}{Y_{\el\ell_2^\prime m_2^\prime}}
\newcommand{\Ylmsix}{Y_{\el\ell_3^\prime m_3^\prime}}

\newcommand{\comm}[1]{\textbf{\textcolor{rossos}{#1}}}
\newcommand{\lsim}{\,\raisebox{-.1ex}{$_{\textstyle <}\atop^{\textstyle\sim}$}\,}
\newcommand{\gsim}{\,\raisebox{-.3ex}{$_{\textstyle >}\atop^{\textstyle\sim}$}\,}

\newcommand{\jl}{j_\ell(k r)}
\newcommand{\jlfourone}{j_{\el\ell_1^\prime}(k_1 r)}
\newcommand{\jlfivetwo}{j_{\el\ell_2^\prime}(k_2 r)}
\newcommand{\jlsixthree}{j_{\el\ell_3^\prime}(k_3 r)}
\newcommand{\jlsix}{j_{\el\ell_3^\prime}(k r)}
\newcommand{\jlthree}{j_{\el\ell_3}(k_3 r)}
\newcommand{\jlthreetau}{j_{\el\ell_3}(k r)}

\newcommand{\Gaunt}{\mathcal{G}_{\el\ell_1^\prime \, \el\ell_2^\prime \, 
\el\ell_3^\prime}^{m_1^\prime m_2^\prime m_3^\prime}}
\newcommand{\Gaunttwo}{\mathcal{G}_{\el\ell_1^\prime \, \el\ell_2^\prime \, 
\el\ell_3}^{m_1^\prime m_2^\prime m_3}}
\newcommand{\Gauntstardef}{\mathcal{H}_{\el\ell_1 \, \el\ell_2 \, \el\ell_3}^{m_1 m_2 m_3}}
\newcommand{\Gauntstarone}{\mathcal{G}_{\el\ell_1 \, L \,\, \el\ell_1^\prime}
^{-m_1 M m_1^\prime}}
\newcommand{\Gauntstartwo}{\mathcal{G}_{\el\ell_2^\prime \, \el\ell_2 \, L}
^{-m_2^\prime m_2 M}}

\newcommand{\de}{{\rm d}}

\newcommand{\dangn}{d \angn}
\newcommand{\dangk}{d \angk}
\newcommand{\dangkone}{d \angk_1}
\newcommand{\dangktwo}{d \angk_2}
\newcommand{\dangkthree}{d \angk_3}
\newcommand{\dk}{d^3 k}
\newcommand{\dkone}{d^3 k_1}
\newcommand{\dktwo}{d^3 k_2}
\newcommand{\dkthree}{d^3 k_3}
\newcommand{\dkfour}{d^3 k_4}
\newcommand{\dallk}{\dkone \dktwo \dk}

\newcommand{\FT}{ \int  \! \frac{d^3k}{(2\pi)^3} 
e^{i\mathbf{k} \cdot \angn \eta_0}}
\newcommand{\planewave}{e^{i\mathbf{k \cdot x}}}
\newcommand{\dallkfourier}{\frac{\dkone}{(2\pi)^3}\frac{\dktwo}{(2\pi)^3}
\frac{\dkthree}{(2\pi)^3}}

\newcommand{\Bis}{B_{\el\ell_1 \el\ell_2 \el\ell_3}^{m_1 m_2 m_3}}
\newcommand{\Avbis}{B_{\el\ell_1 \el\ell_2 \el\ell_3}}

\newcommand{\los}{\mathcal{L}_{\el\ell_3 \el\ell_1 \el\ell_2}^{L \, 
\el\ell_1^\prime \el\ell_2^\prime}(r)}
\newcommand{\loszero}{\mathcal{L}_{\el\ell_3 \el\ell_1 \el\ell_2}^{0 \, 
\el\ell_1^\prime \el\ell_2^\prime}(r)}
\newcommand{\losone}{\mathcal{L}_{\el\ell_3 \el\ell_1 \el\ell_2}^{1 \, 
\el\ell_1^\prime \el\ell_2^\prime}(r)}
\newcommand{\lostwo}{\mathcal{L}_{\el\ell_3 \el\ell_1 \el\ell_2}^{2 \, 
\el\ell_1^\prime \el\ell_2^\prime}(r)}
\newcommand{\losfNL}{\mathcal{L}_{\el\ell_3 \el\ell_1 \el\ell_2}^{0 \, 
\el\ell_1 \el\ell_2}(r)}



\def\d{d}
\def\C{{\rm CDM}}
\def\me{m_e}
\def\te{T_e}
\def\ti{\tau_{\rm initial}}
\def\tci#1{n_e(#1) \sigma_T a(#1)}
\def\tr{\eta_r}
\def\dtr{\delta\eta_r}
\def\dd{\widetilde\Delta^{\rm Doppler}}
\def\dsw{\Delta^{\rm Sachs-Wolfe}}
\def\clsw{C_\ell^{\rm Sachs-Wolfe}}
\def\cldop{C_\ell^{\rm Doppler}}
\def\Dt{\widetilde{\Delta}}
\def\mut{\mu}
\def\vt{\widetilde v}
\def\hp{ {\bf \hat p}}
\def\sdv{S_{\delta v}}
\def\svv{S_{vv}}
\def\bvt{\widetilde{\bv}}
\def\delt{\widetilde{\delta_e}}
\def\cos{{\rm cos}}
\def\nn{\nonumber \\}
\def\bq{ {\bf q} }
\def\ba{ {\bf p} }
\def\bap{ {\bf p'} }
\def\bqp{ {\bf q'} }
\def\bp{ {\bf p} }
\def\bpp{ {\bf p'} }
\def\bk{ {\bf k} }
\def\bx{ {\bf x} }
\def\bv{ {\bf v} }
\def\qp{ p^{\mu}k_{\mu} }
\def\qpp { p^{\mu} k'_{\mu} }
\def\bgm{ {\bf \gamma} }
\def\bkp{ {\bf k'} }
\def\gq{ g(\bq)}
\def\gqp{ g(\bqp)}
\def\fp{ f(\bp)}
\def\h#1{ {\bf \hat #1}}
\def\fpp{ f(\bpp)}
\def\fz{f^{(\vec{0})}(p)}
\def\fpz{f^{(\vec{0})}(p')}
\def\f#1{f^{(#1)}(\bp)}
\def\fps#1{f^{(#1)}(\bpp)}
\def\dq{ {d^3\bq \over (2\pi)^32E(\bq)} }
\def\dqp{ {d^3\bqp \over (2\pi)^32E(\bqp)} }
\def\dpp{ {d^3\bpp \over (2\pi)^32E(\bpp)} }
\def\dtq{ {d^3\bq \over (2\pi)^3} }
\def\dtqp{ {d^3\bqp \over (2\pi)^3} }
\def\dtpp{ {d^3\bpp \over (2\pi)^3} }
\def\part#1;#2 {\partial#1 \over \partial#2}
\def\deriv#1;#2 {d#1 \over d#2}
\def\Done{\Delta^{(1)}}
\def\Dtwo{\widetilde\Delta^{(2)}}
\def\fone{f^{(1)}}
\def\ftwo{f^{(2)}}
\def\tg{T_\gamma}
\def\delpp{\delta(p-p')}
\def\delb{\delta_B}
\def\tc{\eta_0}
\def\DD{\langle|\Delta(k,\mu,\eta_0)|^2\rangle}
\def\DDL{\langle|\Delta(k=l/\tc,\mu)|^2\rangle}
\def\bkpp{{\bf k''}}
\def\kmkp{|\bk-\bkp|}
\def\kmkpsq{k^2+k'^2-2kk'x}
\def\tt{ \left({\tau' \over \tau_c}\right)}
\def\kt{ k\mu \tau_c}

%
%
%


\section{Introduction}
\noindent

Since the invention of supersymmetry, 
 the question of determing  the supersymmetric theory 
that describes the Standard Model (SM) interactions has been at the 
forefront of High Energy Physics. Strong evidence of a new particle found at LHC, the Higgs boson, 
has renewed interest, since, the mass of this particle and its couplings to the rest SM particles will reveal 
where new physics might be hidden \cite{Dine:2007xi}. 
Supersymmetric extensions of the SM, have, among others, the potential to stabilize the weak scale, 
to allow gauge coupling unification,
to provide dark matter candidates and to dynamically explain the hierarchy of weak and Planck scale. In fact,
 it is difficult to imagine a candidate better than supersymmetry for the physics beyond the SM 
in the case of a fundamental Higgs particle.

In the Minimal extension of the SM (MSSM), the Higgs sector is composed of a pair of multiplets $H_u$ and $H_d$.
It is by now a common belief that  any supersymmetric extension of the Standard Model  will necessarily include both 
Higgs fields.
The reason is twofold: first two Higgs fields are required in order to give masses to up- and down quarks as 
holomorphicity of the superpotential does not allow appropriate Yukawa couplings for giving mass to both up- and down-type quarks 
by a single Higgs superfield. Second, simple anomaly arguments lead to an additional Higgs multiplet if quarks and leptons
are organized in usual families. Therefore, either one  considers exact supersymmetry with two Higgs multiplets, or, alternatively
he gets rid of the down-type Higgs for example, at the cost of introducing hard supersymmetry breaking terms (arising basically from
the non-holomorphicity of the superpotential) \cite{Ibe:2010ig}. A difficulty with a chiral Higgs 
sector is that in the absence of $H_d$, the Higgsino is massless until electroweak symmetry breaking. Moreover,
the cancellation of anomalies, previously canceled by $H_d$, requires the introduction of many new fields in various
representations. These new fields should be chiral as well as heavy enough so that they do not 
mess low energy phenomenology.
This is also the case in models with  two Higgs fields and exact supersymmetry, where $H_d$  is just a spectator with no vev and no
coupling to fermions \cite{Davies:2011mp}. Such models, although challenging from the model building point of view,
 have a variety of new fields, which are needed to be introduced in order to 
take over the role of $H_d$, making the models less appealing.

When now  gravity is taken into account, supersymmetry turns out to be a local symmetry with corresponding  
gauge field no other than the gravitino, 
a spin-$\frac{3}{2}$ massless Majorana fermion \cite{deser-zumino,Freedman:1976xh}. If supersymmetry is a fundamental 
symmetry of nature,   it should be broken. In fact the spontaneous breaking of the ${\cal{N}}=1$ supersymmetry 
 implies the existence of a pseudo-Goldstone fermion, the goldstino. The latter will serve as the longitudinal component of 
the  gravitino when local SUSY is broken
 \cite{Deser:1977uq}. This is the super-Higgs mechanism 
which gives mass to gravitino. In a linearly realized supersymmetry, the superpartner of the goldstino is a complex scalar,
the sgoldstino. As it is not protected by any symmetry, it gets a mass. If this mass  is much larger 
than an energy scale, it can be integrated out. In this case, the 
spin-$\frac{3}{2}$ components of the gravitino are highly suppressed, and the phenomenological interesting part is the 
spin-$\frac{1}{2}$ \cite{Fayet:1977vd,Casalbuoni:1988kv,Casalbuoni:1988qd}, namely the goldstino, which possesses  non-linear supersymmetry 
\cite{Volkov:1973ix,Ivanov:1978mx,Rocek:1978nb,Lindstrom:1979kq,uematsu,ho-kim,Casalbuoni:1988xh,luty,lee,brignole,antoniadis1,antoniadis2,bagger,seiberg1}. 
In the opposite case of a light sgoldstino, the latter should be included in the low-energy effective theory.

There exist various formulations for goldstino couplings and non-linear supersymmetry.
Among them, an interesting  framework to discuss non-linear supersymmetry is the  constrained superfield formalism
 \cite{Komargodski:2009rz}. We will consider couplings of the non-linear 
goldstino sector to the MSSM with the use of higher dimensional superspace operators. 
In fact, these couplings of the goldstino to the MSSM
have been
computed  by Antoniadis\,{\it et.\,all}  in a series of papers 
\cite{pantelis,Antoniadis:2012zz} (see also \cite{Antoniadis:2010nb,Antoniadis:2009rn,Antoniadis:2008es} for higher 
dimensional effective operators in the MSSM).
In the  constrained superfield formulation we will employ here, 
we will assume that supersymmetry is spontaneously broken at a SUSY breaking scale $\sqrt{f}$, which will be taken to be
at the multi-{\rm{TeV}} region. Then at energy scales above $\sqrt{f}$, we have MSSM and the goldstino superfield.
At lower scales below SUSY breaking scale $\sqrt{f}$ but above $m_{soft}$ we have again MSSM but the goldstino now is non-linear 
(in the sense that supersymmetry transformations on goldstino are non-linear). Then at low energies below 
$m_{soft}$ only the goldstino fermion couples to SM fields. 
Here, we will discuss energy regions around $m_{soft}$ and below $\sqrt{f}$ where supersymmetry is non-linearly realized on the 
goldstino mode. We will see how the latter can be implemented such that to reduce the Higgs sector in non-linear MSSM. 
%
%

As far as the mass generating mechanism for quarks (and appropriately for leptons) is concerned, 
the   Yukawa couplings of $H_d$ 
\be
 \int d^2 \theta   \bar{d} Q\cdot H_d
\ee
are not available any more. In the models we will present here, mass generation is achieved by 
employing the constrained superfield $X$ and the single Higgs  $H_u$ through the interaction
\be
 \frac{m_{soft}}{f\, \Lambda} \int d^2 \theta d^2  \bar{\theta} \bar{X} \bar{H}_u e^V Q \bar{d}.
\ee
The above coupling emerges from the coupling of the MSSM fields  to the goldstino superfield (suppressed by the cutoff $\Lambda$)
and originates from the replacement
of the spurion $Y\to (m_{soft}/f) X$, where $Y $ is the spurion $Y=\theta^2 m_{soft}$ and $m_{soft}$ is a generic soft mass.
For more details
on this one may consult \cite{Komargodski:2009rz}. 
In particular,  we will present consistent non-linear supersymmetric extensions to the SM that involve:
\begin{itemize}
\item A single  Higgs field $H_u$ where the second Higgs $H_d$ has been replaced by a half family, and
 \item A standard  Higgs $H_u$ where the second Higgs $H_d$ has been turned into a spectator.
\end{itemize}
We note that  in these  SUSY extensions of the SM there is no 
 $\mu$-problem due  to  symmetries or to the spectrum of the theory.

\section{Non-Linear MSSM}

By coupling the non-linear constrained superfield $X$ to the  MSSM  \cite{Komargodski:2009rz}, we get 
 the ``non-linear MSSM'', details of which  has been worked out in \cite{pantelis}. 
Here we will briefly recall its basic features. 
The chiral superfields spectrum of (the two-Higgs) non-linear MSSM is summarized in the following table
\begin{displaymath}
\begin{array}{llll}  \hline \hline 
 \ \   & \text{spin}\ 0 & \text{spin}\ 1/2 & SU(3)_c,\ SU(2)_L,\ U(1)_Y  \\ \hline
Q\ (\times 3) & (\tilde{u}_L, \tilde{d}_L) & (u_L, d_L) &\ \ \ \ \ \ \ \ \mathbf{3},\ \mathbf{2},\ 1/3 \\ 
\bar{u}\ (\times 3) & \tilde{\bar{u}}_L & \bar{u}_L &\ \ \ \ \ \ \ \ \bar{\mathbf{3}},\ \mathbf{1},\ -4/3 \\ 
\bar{d}\ (\times 3) & \tilde{\bar{d}}_L & \bar{d}_L & \ \ \ \ \ \ \ \ \bar{\mathbf{3}},\ \mathbf{1},\ 2/3 \\ 
L\ (\times 3) & (\tilde{{\nu}}_{eL}, \tilde{e}_{L}) & ({\nu}_{eL}, e_{L}) & \ \ \ \ \ \ \ \ \mathbf{1},\ \mathbf{2},\ -1 \\ 
\bar{e}\ (\times 3) & \tilde{\bar{e}}_L & \bar{e}_L & \ \ \ \ \ \ \ \ \mathbf{1},\ \mathbf{1},\ 2 \\ 
H_u & (H_{u}^{+} ,H_{u}^{0}) & (\tilde{H}_{u}^{+} ,\tilde{H}_{u}^{0}) & \ \ \ \ \ \ \ \ \mathbf{1},\ \mathbf{2},\ 1 \\ 
H_d & (H_{d}^{0} ,H_{d}^{-}) & (\tilde{H}_{d}^{0} ,\tilde{H}_{d}^{-}) & \ \ \ \ \ \ \ \ \mathbf{1},\ \mathbf{2},\ -1 \\ 
X  & \phi & G & \ \ \ \ \ \ \ \ \mathbf{1},\ \mathbf{1},\ 0 \\ 
\hline \hline 
\end{array}
\end{displaymath}
\vspace{-.5cm}
\captionof{table}{MSSM chiral superfields spectrum}
\noindent
The theory  is described by the superspace Lagrangian\footnote{Our superspace conventions are those of Wess and Bagger \cite{Wess:1992cp}.}
\be
\label{total2H}
{\cal{L}} = {\cal{L}}_0+{\cal{L}}_g
\ee
where 
\be
\label{total2H2}
{\cal{L}}_0=
{\cal{L}}_{K} + {\cal{L}}_{Yu} + {\cal{L}}_{Yd} + + {\cal{L}}_{\mu} \label{langmssm}
\ee
is the  MSSM superspace Lagrangian 
and 
\be
\label{total2H3}
{\cal{L}}_g= {\cal{L}}_{X} + {\cal{L}}_{s}
+ {\cal{L}}_{tu} + {\cal{L}}_{td} + {\cal{L}}_{B}\, \label{langX}
\ee
describes collectively all the dynamics of the constrained superfield $X$. 
Note that ${\cal{L}}_g$  contains higher dimensional operators 
and hence it is defined with  a cut off \cite{pantelis,Komargodski:2009rz}.
The  Lagrangian (\ref{langmssm}) contains the kinematic terms ${\cal{L}}_K$, Yukawa couplings 
${\cal{L}}_{Y_u},{\cal{L}}_{Y_d}$ as well as the $\mu$- and B-terms ${\cal{L}}_\mu$ and ${\cal{L}}_B$, respectively.  
In particular we have, in standard notation, 
the superspace form \cite{Aitchison:2007fn,Martin:1997ns}
\be
\label{K-superspace}
{\cal{L}}_{K}=
\sum_{\Phi} \int d^4 \theta \bar{\Phi} e^V \Phi + \int d^4 \theta \bar{H}_{d} e^V H_{d}+ 
\int d^4 \theta \bar{H}_{u} e^V H_{u} +
\left\{ \sum_{\rm gauge} \frac{1}{16 g^2 \kappa} \int d^2 \theta \text{Tr}W^{\alpha} W_{\alpha}
+h.c.\right\}
\ee
where $\Phi= Q_i, \bar{u}_i, \bar{d}_i, L_i, \bar{e}_i $, denotes collectively  
the usual quark and lepton chiral superfields with $i=1,...3$ enumerating the three families.
In the gauge sector, 
the  
sum is over  the gauge group of the SM while 
$\kappa$ is a constant to cancel the trace factor. 
The Yukawa couplings are described in superspace as  
\be
\label{Yu}
{\cal{L}}_{Yu}= \int d^2 \theta y^{ij}_{u} \bar{u}_{i} Q_{j} \cdot H_u +h.c.
\ee
and
\be
\label{Yd} 
{\cal{L}}_{Yd}= \int d^2 \theta \Big{(} - y^{ij}_{d} \bar{d}_{i} Q_{j} \cdot H_d  - y^{ij}_{e} \bar{e}_{i} L_{j} \cdot H_d  
\Big{)} +h.c. 
\ee
where $y^{ij}_s,~(s\!=\!e,u,d)$  are the Yukawa matrices of the SM. The dot symbol above 
refers to the SU(2) invariant product of two doublets 
\footnote{For example, if $A$ and $B$ are two $SU(2)$ doublets, $A \cdot B = \epsilon^{ij}A_i B_j $.}.
Finally, the last term of ${\cal{L}}_0$ is the $\mu$-term, which describes a pure interaction 
between the two Higgses
\be
\label{mu}
{\cal{L}}_{\mu}= \mu \int d^2 \theta H_u \cdot H_d +h.c.
\ee
Note that ${\cal{L}}_\mu$ involves the  new parameter $\mu$ which does not have an analog in  
SM theory and  no obvious origin. This term always appears  even if it excluded at tree level as it will
 emerge through quantum corrections, except if a symmetry forbids it.

The constrained superfield (goldstino) Lagrangian has also various contributions. 
The first contribution ${\cal{L}}_g$ has the usual form \cite{Komargodski:2009rz}
\be
\label{XNL}
{\cal{L}}_{X}= \int d^4 \theta \bar{X} X + \left\{ \int d^2 \theta f X  +h.c. \right\}
\ee
with $\sqrt{f}$ the hidden sector SUSY breaking scale. The superfield $X$ satisfies the constraint 
\be
X^2=0,
\ee
and more on this can be found in the appendix.
Soft masses are produced by the following Lagrangian \cite{pantelis}
\be
\label{soft-superspace}
\nonumber
{\cal{L}}_{s} &=& \int d^4 \theta \bar{X} X 
\Big{(} c_{H_u} \bar{H_u} e^V H_u + c_{H_d} \bar{H_d} e^V H_d\Big{)} + 
\sum_{\Phi} c_{\Phi} \int d^4 \theta \bar{X} X \bar{\Phi} e^V \Phi 
\\
&-& \left( \sum_{gauge} \frac{1}{16 g^2 \kappa} \frac{2 m_{\lambda}}{f} \int d^2 \theta X 
\,\text{Tr}W^{\alpha} W_{\alpha} +h.c. \right)
\ee
where
 \be
c_{H_{u,d}}=-\frac{m^{2}_{H_{u,d}}}{f^2}\, , ~~~c_{\Phi}=-\frac{m^{2}_{\Phi}}{f^2}.
\ee
Moreover, the triple scalar coupling terms are given below in superspace form \cite{pantelis,Aitchison:2007fn}
\be
\label{tu}
{\cal{L}}_{tu}= \frac{a_{u}^{ij}}{f} \int d^2 \theta X \bar{u}_{i} Q_{j} \cdot H_u  +h.c.
\ee
and
\be
\label{td}
{\cal{L}}_{td}= - \frac{a_{d}^{ij}}{f} \int d^2 \theta X  \bar{d}_{i} Q_{j} \cdot H_d - \frac{a_{e}^{ij}}{f} \int d^2 \theta X  \bar{e}_{i} L_{j} \cdot H_d +h.c.
\ee
The dimensionfull constants $a_{u}^{ij}, a_{d}^{ij}, a_{e}^{ij}$ are usually taken to be  
\be
\label{A0}
a_{u}^{ij}= A_0 \ y_{u}^{ij},~~~ a_{d}^{ij}= A_0 \ y_{d}^{ij},~~~ a_{e}^{ij}=A_0 \ y_{e}^{ij}
\ee
where  $A_0$ is a mass parameter. 
 The final contribution to ${\cal{L}}_g$ is the  B-term
\be
\label{B}
{\cal{L}}_{B}= \frac{B}{f} \int d^2 \theta X H_u \cdot H_d +h.c.
\ee

We may  proceed by integrating out the auxiliary fields, 
and in particular the auxiliary field of the constrained superfield $X$, 
which we will call it  F. The resulting theory  is the non-linear MSSM. 
Of course, to solve the equations of motion for F, an  expansion in powers of the hidden sector SUSY breaking scale $f$
is needed. 
The full Higgs potential then reads \cite{pantelis}
\be
\label{Two-Higgs-Potential}
{\cal{V}} &=& f^2 + (|\mu| + m^{2}_{u} ) |H_u|^2 + (|\mu| + m^{2}_{d} ) |H_d|^2 + ( B H_u \cdot H_d + h.c. ) \\ 
\nonumber
&+&\frac{1}{f^2} | m^{2}_{u}  |H_u|^2 + m^{2}_{d}  |H_d|^2 +  B H_u \cdot H_d  |^2 + \frac{g^{2}_{1}+g^{2}_{2}}{8} [|H_u|^2 - |H_d|^2 ] +\frac{g^{2}_{2}}{2} |H_{u}^{\dagger} H_d|^2 + {\cal{O}}(\frac{1}{f^3}).
\ee
One exceptional property of any supersymmetric extension of the SM is that it can actually be used to make predictions for 
the Higgs mass. Given  $M_W$, due to supersymmetry, the otherwise free 
Higgs self-coupling $\lambda$  
is now related to the $U(1)$ and  $SU(2)$  couplings $g_1,g_2$ by the relation 
$\lambda \sim g_1^{2} +g_2^{2}$ as can be seen from (\ref{Two-Higgs-Potential}).
Note that the Yukawa couplings in this theory are the same as in the MSSM
\be
\label{Two-Higgs-Yukawa-Couplings}
\nonumber
{\cal{L}}_{Yukawa} = &-& y^{ij}_{u} \bar{u}_{Li}^{ \alpha} (  u_{Lj\alpha} , d_{Lj\alpha}  ) \left( \begin{array}{c} H^{0}_{u} \\ -H^{+}_{u} \end{array} \right)
\\
\nonumber
&+& y^{ij}_{d} \bar{d}_{Li}^{ \alpha} (  u_{Lj\alpha} , d_{Lj\alpha}  ) \left( \begin{array}{c} H^{-}_{d} \\ -H^{0}_{d} \end{array} \right)
\\
&+& y^{ij}_{e} \bar{e}_{Li}^{ \alpha} (  {\nu_{e}}_{Lj\alpha} , e_{Lj\alpha}  ) \left( \begin{array}{c} H^{-}_{d} \\ -H^{0}_{d} \end{array} \right) + h.c.
\ee

\section{Non-Linear MSSM$\bf{3\frac{1}{2}}$ }

Let us recall at this point the two basic reasons for which a  second Higgs field is needed in MSSM 
and in fact in most (if not all)
of the supersymmetric extensions of the SM: 
\begin{enumerate}
\nonumber
\item A second Higgs is needed to cancel the gauge anomaly introduced by a single Higgs supermultiplet.
\item Due to the holomorphicity of the superpotential, a second Higgs is necessary  
in order to write down  Yukawa couplings and give masses to those fermions
the first Higgs cannot.
\end{enumerate}
Therefore, a theory with a single Higgs should be anomaly free and give masses to fermions. 
Mass generation by Yukawa couplings is crucial but before discussing this issue, 
we should make sure that the theory with a single Higgs makes sense, i.e., 
it is anomaly free.
 Therefore, the chiral spectrum should be such so there is no gauge anomaly. 
Anomaly cancellation can be achieved with an additional new ``half-family`` and deviate from standard MSSM. 
The resulting  MSSM$3\frac{1}{2}$ deviations we will present here  are presented in the following table
\begin{table}[!h]
\centering
\renewcommand{\arraystretch}{1.3}
\begin{tabular}{|c|c|}
\hline\hline
\text{Higgs Multiplet:}   & \text{Replaced with:}\\
\hline\hline
\multirow{2}{*}{$H_u$}& $Q,\ \bar{u},\ \bar{d},\ S $\ (\text{or}) \\
\cline{2-2}    
&$\bar{L},\ e,\ S$ \\
\hline\multicolumn{2}{|c|}{or} \\
\hline
\multirow{2}{*}{$H_d$}& $\bar{Q},\ u,\ d,\ \bar{S} $\ (\text{or}) \\
\cline{2-2}    
&$L,\ \bar e,\ \bar S$ \\
\hline\hline
\end{tabular}

\caption{Possible Higgs superfields replacements}
\label{table1}
\end{table}

\noindent
where $S$ is a superfield that has the quantum numbers of $\bar{e}$ but  no
lepton number and it is necessary for anomaly cancellation. Here we will focus on the last possibility in the above  
table and replace $H_d$ by a leptonic generation and $S$. 
We can equally adopt a half-family with only a quark generation, at least at the theoretical level,
which,  nevertheless will lead to different phenomenology.

The number of families is constrained by precision electroweak data \cite{langacker}.   
Direct searches by CDF and D0  set strong limits $m_{t'} > 335 \rm{GeV}$ \cite{CDF1} and $m_{b'} > 385
\rm{GeV}$  
at the 95\%
confidence level for a fourth generation of new $t',b'$ quarks.
LHC also puts more severe constraints in direct searches for extra quarks like
short-lived $b'$ quarks in the signature of trileptons and same-sign dileptons.
CMS for example has ruled out $m_{b'} < 611 \rm{GeV}$ at 95\% confidence level  by assuming exclusive decay 
of $b'\to t\, W$ \cite{CMS1}. 
 Similarly,  no excess over the SM expectations has been observed 
in CMS search for pair production of top-like quarks $t'$, excluding a
fourth generation $t'$ quark with a mass $m_{t'}< 557 \rm{GeV}$ \cite{CMS2}. 
 Also for pair production of a bottom-like  new quark $b'$, 
ATLAS collaboration
reported the exclusion at 95\% confidence level of $b'$ quarks with mass $m_{b'} < 400 \rm{GeV}$
decaying via the channel $b'\to Z+b$ \cite{atlas}. 
 
Extra quarks and leptons are also severely constrained by Higgs production at LHC. 
For example, the dominant source of Higgs production is 
 a single Higgs produced by gluon fusion through a
heavy quark loop. The $gg \to h$ production cross section $\sigma(gg\to h)$
 is proportional to  the Higgs to gluon decay width  $\Gamma(h\to gg)$ which is dominated by heavy quarks with the largest 
Yukawa couplings. This decay width is for example increased by a factor of 5 to 6 relative to SM in fourth generation
models \cite{guo,maltoni}.

As far as a fourth generation of leptons is concerned, the LEP reported the lower bound
for new heavy charged lepton $\tau'$, $m_{\tau'} > 100 \rm{GeV}$\cite{LEP}.
Similarly, the Z invisible
width and the assumption of Dirac masses, set  $m_{\nu'} > m_Z/2$  for new heavy stable neutrinos \cite{paschos}. 
On the other hand, if such new neutral leptons are lighter than
half the Higgs boson mass,   a new invisible channel $H \to \nu' \bar{\nu}'$ is open up increasing the total Higgs width and overtakes 
  the
$H\to f\bar{f}$ rates for example  with a significant
branching ratio in the low mass region.

Returning to our model, the  chiral superfields spectrum is
\begin{displaymath}
\begin{array}{llll}  \hline \hline 
 \ \   & \text{spin}\ 0 & \text{spin}\ 1/2 & SU(3)_c,\ SU(2)_L,\ U(1)_Y  \\ \hline
Q\ (\times 3) & (\tilde{u}_L, \tilde{d}_L) & (u_L, d_L) &\ \ \ \ \ \ \ \ \mathbf{3},\ \mathbf{2},\ 1/3 \\ 
\bar{u}\ (\times 3) & \tilde{\bar{u}}_L & \bar{u}_L &\ \ \ \ \ \ \ \ \bar{\mathbf{3}},\ \mathbf{1},\ -4/3 \\ 
\bar{d}\ (\times 3) & \tilde{\bar{d}}_L & \bar{d}_L & \ \ \ \ \ \ \ \ \bar{\mathbf{3}},\ \mathbf{1},\ 2/3 \\ 
L\ (\times 4) & (\tilde{{\nu}}_{eL}, \tilde{e}_{L}) & ({\nu}_{eL}, e_{L}) & \ \ \ \ \ \ \ \ \mathbf{1},\ \mathbf{2},\ -1 \\ 
\bar{e}\ (\times 4) & \tilde{\bar{e}}_L & \bar{e}_L & \ \ \ \ \ \ \ \ \mathbf{1},\ \mathbf{1},\ 2 \\ 
H_u & (H_{u}^{+} ,H_{u}^{0}) & (\tilde{H}_{u}^{+} ,\tilde{H}_{u}^{0}) & \ \ \ \ \ \ \ \ \mathbf{1},\ \mathbf{2},\ 1 \\ 
\bar{S}  & \bar{s} & \tilde{\bar{s}} & \ \ \ \ \ \ \ \ \mathbf{1},\ \mathbf{1},\ -2 \\ 
X  & \phi & G & \ \ \ \ \ \ \ \ \mathbf{1},\ \mathbf{1},\ 0 \\
\hline \hline
\end{array}
\end{displaymath}
\vspace{-.5cm}
\captionof{table}{Single Higgs Non-Linear MSSM$3\frac{1}{2}$}
\noindent


Even if the theory  is anomaly free, 
we are still facing the 
problem of how to give masses to quarks and leptons while maintaining SUSY as the second Higgs $H_d$ is missing. 
For this reason,  we may introduce higher dimensional operators to replace the Yukawa couplings (\ref{Yd}). 
The Lagrangian that will replace ${\cal{L}}_{Yd}$ in (\ref{Yd}) is
\be
\label{Yd'-superspace}
\nonumber
{\cal{L}}_{Yd'} = - \frac{m_{soft}}{f\, \Lambda} \int d^2 \theta d^2 \bar{\theta} \bar{X} 
\Big{(} y^{ij}_{d}  \bar{H}_{u} e^V Q_{j} \bar{d}_{i} + y^{IJ}_{e}  \bar{H}_{u} e^V L_{J} \bar{e}_{I} \Big{)}+ h.c. 
\\
= - \frac{m_{soft}}{16f\, \Lambda} D^2 \bar{D}^2  \bar{X} \Big{( }
 y^{ij}_{d}  \bar{H}_{u} e^V Q_{j} \bar{d}_{i} + y^{IJ}_{e}  \bar{H}_{u} e^V L_{J} \bar{e}_{I} \Big{)} \Big{|} + h.c.
\ee
where now $I,J=1,...4$ run over the fourth lepton generation. We recall again that the factor $m_{soft}/f$ emerges by the replacement 
of the spurion $Y=\theta^2m_{soft}$ by $(m_{soft}/f) X$ as we have pointed out already in the introduction \cite{Komargodski:2009rz}.
In component form (\ref{Yd'-superspace}) turns out to be
\be
\label{Yd'-components}
\nonumber
{\cal{L}}_{Yd'} &=&  
 \frac{m_{soft}}{f\, \Lambda}  \, y^{ij}_{d} \left\{
\phantom{\frac{X^x}{X^x}}\!\!\!\!\!\!\!\!\! \right.\bar{F} ( \bar{H}^{+}_{u}, \bar{H}^{0}_{u}) \left( \begin{array}{c} u_{Lj}^{\alpha} 
\\  d_{Lj}^{\alpha} \end{array} \right) \bar{d}_{Li \alpha} 
-\bar{F} (  \bar{H}^{+}_{u}, \bar{H}^{0}_{u} )  \left( \begin{array}{c} \tilde{u}_{Lj} \\  
\tilde{d}_{Lj} \end{array} \right)  F_{\bar{d}_{Li}}
\\
\nonumber
&&\hspace{1.3cm}
- \left.
\bar{F} (  \bar{H}^{+}_{u}, \bar{H}^{0}_{u})   \left( \begin{array}{c} F_{u_Lj} \\  F_{ d_Lj}  \end{array} \right) 
 \tilde{\bar{d}}_{Li} \right\}
\\
\nonumber
&&+\frac{m_{soft}}{f\, \Lambda}\,  y^{IJ}_{e} \left\{
\phantom{\frac{X^x}{X^x}}\!\!\!\!\!\!\!\!\! \right.
 \bar{F} ( \bar{H}^{+}_{u}, \bar{H}^{0}_{u}) \left( \begin{array}{c} {\nu}_{eLJ}^{\alpha} \\ e_{LJ}^{\alpha} \end{array} \right) 
\bar{e}_{LI \alpha}
-\bar{F} (  \bar{H}^{+}_{u}, \bar{H}^{0}_{u} )  \left( \begin{array}{c} \tilde{\nu}_{eLJ} \\  \tilde{e}_{LJ} \end{array} \right)  
F_{\bar{e}_{LI}}
\\
&&\hspace{1.3cm}
-\left. \bar{F} \big{(}  \bar{H}^{+}_{u}, \bar{H}^{0}_{u}\big{)}   
\left( \begin{array}{c} F_{{\nu}_{eL}J} \\  F_{ e_LJ}  
\end{array} \right)  \tilde{\bar{e}}_{LI} \right\} +h.c.
\ee
where we recall that $F$ is the auxiliary field of the goldstino superfield. 
In the above equation (\ref{Yd'-components}) we have kept  only the terms with  no goldstino couplings. 
In the appendix, the higher dimensional operator that serves as a building block for the full Lagrangian is given
 in terms of the goldstino and its lowest component $\phi$, which is integrated out to obtain the
 non-linear supersymmetric Lagrangian. In this framework a natural explanation of the scale $f$ is 
proposed and our non-renormalizable operators (\ref{Yd'-superspace}) fit well to the general picture 
\cite{Komargodski:2009rz,pantelis,Antoniadis:2012zz}. The Higgs triple scalar couplings Lagrangian 
to replace ${\cal{L}}_{td}$ in (\ref{td}) is, in superspace form
\be
\label{td'-superspace}
\nonumber
{\cal{L}}_{td'} =  &-& \frac{m_{soft}^2}{f^2\, \Lambda^2}  \int d^2 \theta d^2 \bar{\theta}  \bar{X} X  \left\{a^{ij}_{d} 
\bar{H}_{u} e^V Q_{j} \bar{d}_{i} + a^{IJ}_{e}  \bar{H}_{u} e^V L_{J} \bar{e}_{I}  \right\} + h.c. 
\\
= &-& \frac{m_{soft}^2}{16 f^2\, \Lambda^2} D^2 \bar{D}^2  \bar{X} X  \left\{  a^{ij}_{d} \bar{H}_{u} e^V Q_{j} \bar{d}_{i} +
 a^{IJ}_{e}  \bar{H}_{u} e^V L_{J} \bar{e}_{I}   \right\} \Big{|}  +  h.c.
\ee
After performing the superspace integration we get
\be
\label{td'-components}
{\cal{L}}_{td'} =  - \frac{m_{soft}^2}{f^2\, \Lambda^2}   \bar{F} F \left\{    a^{ij}_{d} ( \bar{H}^{+}_{u}, \bar{H}^{0}_{u})  
\left( \begin{array}{c} \tilde{u}_{Lj} \\  \tilde{d}_{Lj} \end{array} \right) \tilde{\bar{d}}_{Li} +  a^{IJ}_{e} 
( \bar{H}^{+}_{u}, \bar{H}^{0}_{u})  \left( \begin{array}{c} \tilde{\nu}_{eLJ} \\  \tilde{e}_{LJ} \end{array} \right)
 \tilde{\bar{e}}_{LI} \right\} +  h.c.
\ee
where goldstino couplings have been ignored. Then, it is clear that the replacements
\be
\label{replace}
\nonumber
&&{\cal{L}}_{Yd} \rightarrow  {\cal{L}}_{Yd'}\\
&&{\cal{L}}_{td} \rightarrow  {\cal{L}}_{td'}
\ee
in (\ref{total2H2}) and (\ref{total2H3}) respectively give rise to (non-linear) MSSM with only one Higgs (the $H_u$). 

We may proceed further and  integrate out the auxiliary sector of the goldstino superfield.
This will uncover the on-shell Lagrangian with  Yukawa  and triple scalar couplings. 
Since in this work we are only interested in the standard model sector, 
we will not write down any goldstino couplings when solving the equations of motion of the auxiliary fields. 
This greatly simplifies the results without spoiling the final answer. Nevertheless it is important 
to study the implications of these new terms that include the goldstino as well, but this is left for future work. 
The relevant terms in our total Lagrangian (\ref{total2H}) are therefore
\be
\allowdisplaybreaks[2]
\label{F1}
\nonumber
{\cal{L}}_{F}&=& \frac{1}{2} F \bar{F} + f \bar{F}  + \frac{1}{2} c_{H_u} \bar{F} F |H_u|^2 + \frac{1}{2}\sum_i c_i \bar{F} F |\tilde{\Phi}_i|^2 
\\
\nonumber
&+& \frac{m_{soft}}{f\, \Lambda}\,   y^{ij}_{d} \bar{F} ( \bar{H}^{+}_{u}, \bar{H}^{0}_{u}) 
\left( \begin{array}{c} u_{Lj}^{\alpha} \\ d_{Lj}^{\alpha} \end{array} \right) \bar{d}_{Li \alpha}
-\frac{m_{soft}}{f\, \Lambda}\,  y^{ij}_{d} \bar{F} (  \bar{H}^{+}_{u}, \bar{H}^{0}_{u} )  \left( \begin{array}{c} \tilde{u}_{Lj} \\  
\tilde{d}_{Lj} \end{array} \right)  F_{\bar{d}_{Li}}
\\
\nonumber
&-& \frac{m_{soft}}{f\, \Lambda}\,  y^{ij}_{d} \bar{F} (  \bar{H}^{+}_{u}, \bar{H}^{0}_{u})   \left( \begin{array}{c} F_{u_Lj} \\  F_{ d_Lj}  \end{array} \right)  \tilde{\bar{d}}_{Li} 
-\frac{m_{soft}}{f\, \Lambda}\,   y^{IJ}_{e} \bar{F} (  \bar{H}^{+}_{u}, \bar{H}^{0}_{u})   \left( \begin{array}{c} F_{{\nu}_{eL}J} \\  F_{ e_LJ}  \end{array} \right)  \tilde{\bar{e}}_{LI}  
\\
\nonumber
&+&  \frac{m_{soft}}{f\, \Lambda}\,  y^{IJ}_{e}  \bar{F} ( \bar{H}^{+}_{u}, \bar{H}^{0}_{u}) \left( \begin{array}{c} {\nu}_{eLJ}^{\alpha} \\ e_{LJ}^{\alpha} \end{array} \right) \bar{e}_{LI \alpha}
-\frac{m_{soft}}{f\, \Lambda}\,   y^{IJ}_{e} \bar{F} (  \bar{H}^{+}_{u}, \bar{H}^{0}_{u} )  \left( \begin{array}{c} \tilde{\nu}_{eLJ} \\  \tilde{e}_{LJ} \end{array} \right)  F_{\bar{e}_{L}I}
\\
\nonumber
&-& \frac{m_{soft}^2}{f^2\, \Lambda^2}\,   \bar{F} F     a^{ij}_{d} ( \bar{H}^{+}_{u}, \bar{H}^{0}_{u}) 
\left( \begin{array}{c} \tilde{u}_{Lj} \\  \tilde{d}_{Lj} \end{array} \right) \tilde{\bar{d}}_{Li} -
\frac{m_{soft}^2}{f^2\, \Lambda^2}\,  \bar{F} F  a^{IJ}_{e} ( \bar{H}^{+}_{u}, \bar{H}^{0}_{u})  \left( \begin{array}{c} \tilde{\nu}_{eLJ} \\  \tilde{e}_{LJ} \end{array} \right) \tilde{\bar{e}}_{LI} 
\\
&+& + \sum_i \frac{m_{{\lambda}_{i}}}{2f} F {\lambda}_{i}^2 + h.c.
\ee
where by $\tilde{\Phi}_i$ we denote the lowest components of the various chiral superfields (the sparticles in our case). 
Assuming that  $f$ is large, we may use the expansion
\be
\label{expansion}
\nonumber
\hspace{-.5cm}&& \left(1-\frac{m_{H_{u}}^{2}}{f^2}|H_{u}|^2-\frac{m_{\Phi_i}^{2}}{f^2}|\tilde{\Phi}_i|^2 
- \frac{m_{soft}^2}{f^2\, \Lambda^2}\,  a^{ij}_{d} ( \bar{H}^{+}_{u}, \bar{H}^{0}_{u}) 
\left( \begin{array}{c} \tilde{u}_{Lj} \\  \tilde{d}_{Lj} \end{array} \right) \tilde{\bar{d}}_{Li} 
- \frac{m_{soft}^2}{f^2\, \Lambda^2}\,  a^{IJ}_{e} ( \bar{H}^{+}_{u}, \bar{H}^{0}_{u})  \left( \begin{array}{c} \tilde{\nu}_{eLJ} \\  \tilde{e}_{LJ} \end{array} \right) \tilde{\bar{e}}_{LI} \right)^{-1}
\\
\hspace{-.5cm}&&\simeq 1+\frac{m_{H_{u}}^{2}}{f^2}|H_{u}|^2+\frac{m_{\Phi_i}^{2}}{f^2}|\tilde{\Phi}_i|^2 + 
\frac{m_{soft}^2}{f^2\, \Lambda^2}\, a^{ij}_{d} ( \bar{H}^{+}_{u}, \bar{H}^{0}_{u})  \left( \begin{array}{c} \tilde{u}_{Lj} \\  \tilde{d}_{Lj} \end{array} \right) \tilde{\bar{d}}_{Li} 
+ \frac{m_{soft}^2}{f^2\, \Lambda^2}\,  a^{IJ}_{e} ( \bar{H}^{+}_{u}, \bar{H}^{0}_{u})  \left( \begin{array}{c} \tilde{\nu}_{eLJ} \\  \tilde{e}_{LJ} \end{array} \right) \tilde{\bar{e}}_{LI} .
\nonumber
\ee
in order to eliminate $F$ from (\ref{F1}) so that
\be
\allowdisplaybreaks[2]
\label{F1-on-shell}
\nonumber
{\cal{L}}_{{\rm{F,on-shell}}}= &-& \frac{1}{2} f^2  - \frac{1}{2} m_{H_u}^{2} |H_u|^2 - \frac{1}{2} m_{\Phi_i}^{2} |\tilde{\Phi}_i|^2 
\\
\nonumber
&-&   \frac{m_{soft}}{ \Lambda}\,  y^{ij}_{d}  ( \bar{H}^{+}_{u}, \bar{H}^{0}_{u}) \left( \begin{array}{c} u_{Lj}^{\alpha} 
\\ d_{Lj}^{\alpha} \end{array} \right) \bar{d}_{Li \alpha}
+ \frac{m_{soft}}{ \Lambda}\,  y^{ij}_{d}  (  \bar{H}^{+}_{u}, \bar{H}^{0}_{u} )  \left( \begin{array}{c} \tilde{u}_{Lj} \\  \tilde{d}_{Lj} \end{array} \right)  
F_{\bar{d}_{Li}}
\\
\nonumber
&+& \frac{m_{soft}}{\Lambda}\,  y^{ij}_{d}  (  \bar{H}^{+}_{u}, \bar{H}^{0}_{u})   \left( \begin{array}{c} F_{u_Lj} \\  F_{ d_Lj}  \end{array} \right)  \tilde{\bar{d}}_{Li} 
+  \frac{m_{soft}}{ \Lambda}\, y^{IJ}_{e}  (  \bar{H}^{+}_{u}, \bar{H}^{0}_{u})   \left( \begin{array}{c} F_{{\nu}_{eL}J} \\  F_{ e_LJ}  \end{array} \right)  \tilde{\bar{e}}_{LI}  
\\
\nonumber
&-& \frac{m_{soft}}{\Lambda}\,  y^{IJ}_{e}  ( \bar{H}^{+}_{u}, \bar{H}^{0}_{u}) \left( \begin{array}{c} {\nu}_{eLJ}^{\alpha} \\ e_{LJ}^{\alpha} \end{array} \right) \bar{e}_{LI \alpha}
+  \frac{m_{soft}}{\Lambda}\, y^{IJ}_{e}  (  \bar{H}^{+}_{u}, \bar{H}^{0}_{u} )  \left( \begin{array}{c} \tilde{\nu}_{eLJ} \\  \tilde{e}_{LJ} \end{array} \right)  F_{\bar{e}_{L}I}
\\
\nonumber
&-&   \frac{m_{soft}^2}{ \Lambda^2}\,  a^{ij}_{d} ( \bar{H}^{+}_{u}, \bar{H}^{0}_{u})  \left( \begin{array}{c} \tilde{u}_{Lj} \\  \tilde{d}_{Lj} \end{array} \right) \tilde{\bar{d}}_{Li} 
-   \frac{m_{soft}^2}{ \Lambda^2}\, a^{IJ}_{e} ( \bar{H}^{+}_{u}, \bar{H}^{0}_{u})  \left( \begin{array}{c} \tilde{\nu}_{eLJ} \\  \tilde{e}_{LJ} \end{array} \right) \tilde{\bar{e}}_{LI} 
\\
&-& \frac{1}{2} m_{\lambda_i} \lambda^{2}_{i} + h.c. + {\cal{O}}(\frac{1}{f^2}).
\ee
Note that the larger the SUSY breaking scale the better the 
approximation. For a smaller SUSY breaking scale one has to include higher orders in the $\frac{1}{f}$ 
expansion, which leads to new interesting results as in the two-Higgs scenario \cite{pantelis}. 


Therefore, the Yukawa couplings in our theory (\ref{total2H}) with the replacements (\ref{replace}) are
\be
\label{One-Higgs-Yukawa-Couplings}
\nonumber
{\cal{L}}_{Yukawa} = &-& y^{ij}_{u} \bar{u}_{Li}^{ \alpha} (  u_{Lj\alpha} , d_{Lj\alpha}  ) \left( \begin{array}{c} H^{0}_{u} \\ -H^{+}_{u} \end{array} \right)
\\
\nonumber
&-&  \frac{m_{soft}}{\Lambda}\,  y^{ij}_{d}  ( \bar{H}^{+}_{u}, \bar{H}^{0}_{u}) \left( \begin{array}{c} u_{Lj}^{\alpha} \\ d_{Lj}^{\alpha} \end{array} \right) \bar{d}_{Li \alpha}
\\
&-& \frac{m_{soft}}{ \Lambda}\, y^{IJ}_{e}  ( \bar{H}^{+}_{u}, \bar{H}^{0}_{u}) \left( \begin{array}{c} {\nu}_{eLJ}^{\alpha} \\ e_{LJ}^{\alpha} \end{array} \right) \bar{e}_{LI \alpha} + h.c.
\ee
Let us note that an interesting hierarchy has emerged. 
Namely, assuming the same order for $y^{ij}_u, ~y^{ij}_d,~ y^{IJ}_e$
we see that the effective Yukawa couplings  for the bottom and tau 
are suppressed by a factor $m_{soft}/\Lambda$. Thus,
the bottom quark  and $\tau$ lepton masses  $m_b$ and $m_\tau$,  respectively, 
should be of the same order and suppressed by  
$m_{soft}/\Lambda$ with respect to the top quark mass $m_t$
\be
m_b\sim m_\tau\sim \frac{m_{soft}}{\Lambda}m_t
\ee
This is 
indeed the case for a cutoff $\Lambda$ of the order $\Lambda\sim 100\,  m_{soft}$. With $\Lambda\sim \sqrt{f}$
we get that 
$\sqrt{f}\sim 100 m_{soft}$, whereas a cutoff $\Lambda\sim f/m_{soft}$
gives rise to the $\sqrt{f}\sim m_{soft}$ estimate\footnote{We would like to thank E. Dudas and P. Tziveloglou for 
pointing this out.}.

The Higgs potential is given by
\be
\label{MSSM-3+1/2-Higgs-Potential}
{\cal{V}} &=& f^2 + m^{2}_{u} |H_u|^2 + \frac{1}{f^2}  m^{4}_{u}  |H_u|^4  + \frac{g^{2}_{1}+g^{2}_{2}}{8}  |H_u|^4  + {\cal{O}}(\frac{1}{f^3}).
\ee
Radiative corrections to the Higgs potential are expected to drive the quadratic term negative and trigger 
electroweak symmetry breaking. Moreover, this effect is strengthened by the extra Yukawa coupling due to the 
new half-family. The explicit calculation of the 1-loop effective potential can place strong upper and lower 
bounds to the new leptonic family mass.
The tree level prediction for the Higgs mass however is
\be
\label{H-mass}
M_{H_u}^{2}=M_{Z}^{2}+\frac{8M_{W}^{2}m_{u}^{4}}{g_{2}^{2}f^2}+{\cal O}(\frac{1}{f^4})
\ee
Thus, as $f\to \infty$ we have $M_{H_u}\to M_Z$. Therefore, for very large SUSY breaking scale $\sqrt{f}$,
the Higgs mass  saturates the MSSM inequality $M_{H_u}\leq M_Z$. This saturation within MSSM corresponds to 
large $\tan\beta$. By adjusting 
$\sqrt{f}$, we may increase the tree-level Higgs mass so that quantum corrections may shift it to the 
measured value of around $126.5 \rm{GeV}$.    
We plot below the dependence of the tree level Higgs mass to the supersymmetry breaking scale $\sqrt f$, for 
the single Higgs models.
\begin{figure}[!h] \centering{
\includegraphics[scale=0.25]{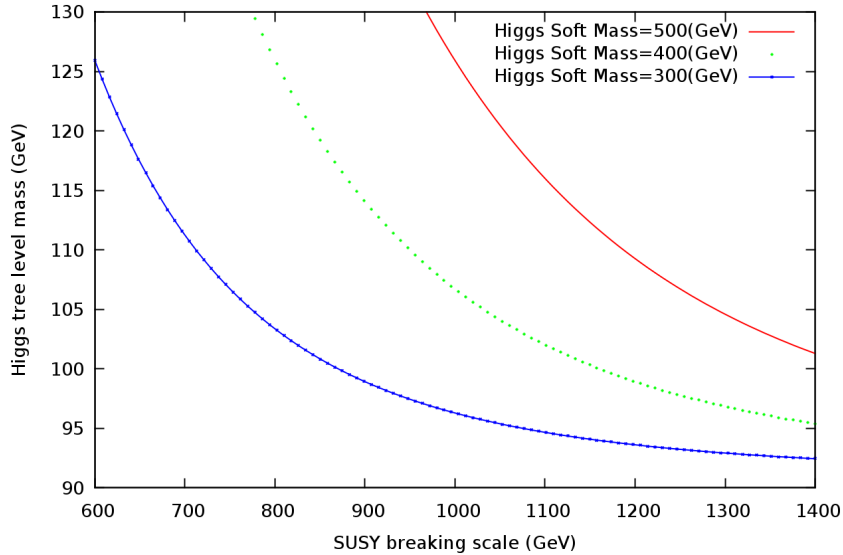}}
\caption{The dependence of the Higgs mass ($M_{H_u}$) on the hidden sector SUSY breaking scale $\sqrt{f}$, with the Higgs soft mass as a parameter.}
\end{figure}

\section{Spectator $H_d$}

As we have seen, Yukawa couplings of $H_d$ can be replaced by the higher dimensional operators of the form
(\ref{Yd'-superspace}) with the help of the constrained superfield $X$. Therefore, we can keep in the spectrum $H_d$
 just to cancel the anomalies 
but use (\ref{Yd'-superspace}) to generate fermion masses. This is possible as long as we can 
avoid couplings of $H_d$ to matter.
This can be achieved by imposing  
 a $\mathbb{Z}_2$ symmetry. 
This symmetry will forbid  interactions like (\ref{mu}) and (\ref{B}). At the same time 
standard MSSM Yukawa couplings (\ref{Yu}) of $H_d$ will not be allowed as well,  again due to the same $Z_2$ symmetry. 
Of course this is different from the case of wrong Higgs couplings of the MSSM where SUSY is hardly broken \cite{Haber:2007dj}. 
The chiral superfields spectrum and its $\mathbb{Z}_2$ assignment  is
\begin{displaymath}
\begin{array}{lllll}  \hline \hline 
 \ \   & \text{spin}\ 0 & \text{spin}\ 1/2 & SU(3)_c,\ SU(2)_L,\ U(1)_Y & Z_2 \\ \hline
Q\ (\times 3) & (\tilde{u}_L, \tilde{d}_L) & (u_L, d_L) &\ \ \ \ \ \ \ \ \mathbf{3},\ \mathbf{2},\ 1/3 & +1 \\ 
\bar{u}\ (\times 3) & \tilde{\bar{u}}_L & \bar{u}_L &\ \ \ \ \ \ \ \ \bar{\mathbf{3}},\ \mathbf{1},\ -4/3 & +1 \\ 
\bar{d}\ (\times 3) & \tilde{\bar{d}}_L & \bar{d}_L & \ \ \ \ \ \ \ \ \bar{\mathbf{3}},\ \mathbf{1},\ 2/3 & +1 \\ 
L\ (\times 3) & (\tilde{{\nu}}_{eL}, \tilde{e}_{L}) & ({\nu}_{eL}, e_{L}) & \ \ \ \ \ \ \ \ \mathbf{1},\ \mathbf{2},\ -1 & +1 \\ 
\bar{e}\ (\times 3) & \tilde{\bar{e}}_L & \bar{e}_L & \ \ \ \ \ \ \ \ \mathbf{1},\ \mathbf{1},\ 2 & +1 \\ 
H_u & (H_{u}^{+} ,H_{u}^{0}) & (\tilde{H}_{u}^{+} ,\tilde{H}_{u}^{0}) & \ \ \ \ \ \ \ \ \mathbf{1},\ \mathbf{2},\ 1 & +1\\ 
H_d & (H_{d}^{0} ,H_{d}^{-}) & (\tilde{H}_{d}^{0} ,\tilde{H}_{d}^{-}) & \ \ \ \ \ \ \ \ \mathbf{1},\ \mathbf{2},\ -1 & -1 \\
X  & \phi & G & \ \ \ \ \ \ \ \ \mathbf{1},\ \mathbf{1},\ 0 & +1 \\
\hline \hline
\end{array}
\end{displaymath}
\vspace{-.5cm}
\captionof{table}{Single-Higgs MSSM with a second spectator Higgs}
\noindent


We see that we keep here the second Higgs just for anomaly cancellation reasons and we have excluded 
any couplings to the MSSM, by imposing a $\mathbb{Z}_2$ symmetry.
In other words, $H_d$ is just a spectator and only $H_u$ has  Yukawa and triple scalar couplings.
The Lagrangian that will take the place of ${\cal{L}}_{Yd}$ in (\ref{Yd}) is then
\be
\label{Yd''-superspace}
\nonumber
{\cal{L}}_{Yd''} = - \frac{m_{soft}}{f\, \Lambda}\, \int d^2 \theta d^2 \bar{\theta} \bar{X} 
\Big{(} y^{ij}_{d}  \bar{H}_{u} e^V Q_{j} \bar{d}_{i} + y^{ij}_{e}  \bar{H}_{u} e^V L_{j} \bar{e}_{i} \Big{)}+ h.c. 
\\
= - \frac{m_{soft}}{16f\, \Lambda}\,D^2 \bar{D}^2  \bar{X} \Big{( }
 y^{ij}_{d}  \bar{H}_{u} e^V Q_{j} \bar{d}_{i} + y^{ij}_{e}  \bar{H}_{u} e^V L_{j} \bar{e}_{i} \Big{)} \Big{|} + h.c.
\ee
which in component form turns out to be
\be
\label{Yd''-components}
\nonumber
{\cal{L}}_{Yd''} &=&  
 \frac{m_{soft}}{f\, \Lambda}\, y^{ij}_{d} \left\{
\phantom{\frac{X^x}{X^x}}\!\!\!\!\!\!\!\!\! \right.\bar{F} ( \bar{H}^{+}_{u}, \bar{H}^{0}_{u}) \left( \begin{array}{c} u_{Lj}^{\alpha} 
\\  d_{Lj}^{\alpha} \end{array} \right) \bar{d}_{Li \alpha} 
-\bar{F} (  \bar{H}^{+}_{u}, \bar{H}^{0}_{u} )  \left( \begin{array}{c} \tilde{u}_{Lj} \\  
\tilde{d}_{Lj} \end{array} \right)  F_{\bar{d}_{Li}}
\\
\nonumber
&&\hspace{1.3cm}
- \left.
\bar{F} (  \bar{H}^{+}_{u}, \bar{H}^{0}_{u})   \left( \begin{array}{c} F_{u_Lj} \\  F_{ d_Lj}  \end{array} \right) 
 \tilde{\bar{d}}_{Li} \right\}
\\
\nonumber
&&+\frac{m_{soft}}{f\, \Lambda}\, y^{ij}_{e} \left\{
\phantom{\frac{X^x}{X^x}}\!\!\!\!\!\!\!\!\! \right.
 \bar{F} ( \bar{H}^{+}_{u}, \bar{H}^{0}_{u}) \left( \begin{array}{c} {\nu}_{eLj}^{\alpha} \\ e_{Lj}^{\alpha} \end{array} \right) 
\bar{e}_{Li \alpha}
-\bar{F} (  \bar{H}^{+}_{u}, \bar{H}^{0}_{u} )  \left( \begin{array}{c} \tilde{\nu}_{eLj} \\  \tilde{e}_{Lj} \end{array} \right)  
F_{\bar{e}_{Li}}
\\
&&\hspace{1.3cm}
-\left. \bar{F} \big{(}  \bar{H}^{+}_{u}, \bar{H}^{0}_{u}\big{)}   
\left( \begin{array}{c} F_{{\nu}_{eL}j} \\  F_{ e_Lj}  
\end{array} \right)  \tilde{\bar{e}}_{Li} \right\} +h.c.
\ee
In the above equation (\ref{Yd''-components}) we have kept  only the terms with  no 
goldstino couplings. The one-Higgs triple scalar couplings Lagrangian 
to replace ${\cal{L}}_{td}$ in (\ref{td}) is, in superspace form
\be
\label{td''-superspace}
\nonumber
{\cal{L}}_{td''} =  &-& \frac{m_{soft}^2}{f^2\, \Lambda^2}\,  \int d^2 \theta d^2 \bar{\theta}  \bar{X} X  \left\{a^{ij}_{d} 
\bar{H}_{u} e^V Q_{j} \bar{d}_{i} + a^{ij}_{e}  \bar{H}_{u} e^V L_{j} \bar{e}_{i}  \right\} + h.c. 
\\
= &-& \frac{m_{soft}^2}{16 f^2\, \Lambda^2}\, D^2 \bar{D}^2  \bar{X} X  \left\{  a^{ij}_{d} \bar{H}_{u} e^V Q_{j} \bar{d}_{i} +
 a^{ij}_{e}  \bar{H}_{u} e^V L_{j} \bar{e}_{i}   \right\} \Big{|}  +  h.c.
\ee
After performing the superspace integration we get
\be
\label{td''-components}
{\cal{L}}_{td''} =  - \frac{m_{soft}^2}{f^2\, \Lambda^2}\,    \bar{F} F \left\{    a^{ij}_{d} ( \bar{H}^{+}_{u}, \bar{H}^{0}_{u})  
\left( \begin{array}{c} \tilde{u}_{Lj} \\  \tilde{d}_{Lj} \end{array} \right) \tilde{\bar{d}}_{Li} +  a^{ij}_{e} 
( \bar{H}^{+}_{u}, \bar{H}^{0}_{u})  \left( \begin{array}{c} \tilde{\nu}_{eLj} \\  \tilde{e}_{Lj} \end{array} \right)
 \tilde{\bar{e}}_{Li} \right\} +  h.c.
\ee
where we have ignored any goldstino couplings. Then, it is clear that  the replacements
\be
\label{replace''}
\nonumber
&&{\cal{L}}_{Yd} \rightarrow  {\cal{L}}_{Yd''}\\
&&{\cal{L}}_{td} \rightarrow  {\cal{L}}_{td''}
\ee
in (\ref{langmssm}) give rise to (non-linear) MSSM with only one Higgs (the $H_u$). For completeness,  
the Higgs potential of this theory is again
\be
\label{One-Higgs-Potential}
{\cal{V}} &=& f^2 + m^{2}_{u} |H_u|^2 + \frac{1}{f^2}  m^{4}_{u}  |H_u|^4  + \frac{g^{2}_{1}+g^{2}_{2}}{8}  |H_u|^4  + {\cal{O}}(\frac{1}{f^3})
\ee
while the Yukawa couplings are
\be
\label{One-One-Higgs-Yukawa-Couplings}
\nonumber
{\cal{L}}_{Yukawa} = &-&y^{ij}_{u} \bar{u}_{Li}^{ \alpha} (  u_{Lj\alpha} , d_{Lj\alpha}  ) \left( \begin{array}{c} H^{0}_{u} \\ -H^{+}_{u} \end{array} \right)
\\
\nonumber
&-&   \frac{m_{soft}}{\Lambda}\,   y^{ij}_{d}  ( \bar{H}^{+}_{u}, \bar{H}^{0}_{u}) \left( \begin{array}{c} u_{Lj}^{\alpha} \\ d_{Lj}^{\alpha} \end{array} \right) \bar{d}_{Li \alpha}
\\
&-&   \frac{m_{soft}}{\Lambda}\, y^{ij}_{e}  
( \bar{H}^{+}_{u}, \bar{H}^{0}_{u}) \left( \begin{array}{c} {\nu}_{eLj}^{\alpha} 
\\ e_{Lj}^{\alpha} \end{array} \right) \bar{e}_{Li \alpha} + h.c.
\ee
Note also that the $\mathbb{Z}_2$ symmetry does not allow $\mu$- and $B$-terms. 

\section{Conclusions }

The main purpose of this work was to show that in the non-linear  MSSM framework,  
a one Higgs doublet 
is possible and equally motivating with the two-Higgs scenario.
In fact, even when dealing with a two-Higgs MSSM, unavoidably, 
non-linear goldstino dynamics should be considered as a possibility for the physics beyond MSSM.
In this context, higher dimensional operators are introduced in order to study the consequences of the non-linearities 
 of the underlying theory.
However,
 higher dimensional operators is what is needed for a single Higgs MSSM.
In this sense, a single Higgs MSSM is quite interesting, as it turns out that it is intrinsically 
connected to the underlying supergravity theory, as it cannot be constructed without the use of 
the higher dimensional operators.

In this approach, we have constructed two consistent supersymmetric extensions of the SM where 
only one scalar field is required to have a non-trivial vacuum expectation value.
The energy regime of both models is comparable or above the soft masses. 
In the first model, the second Higgs superfield is completely missing from the MSSM spectrum 
 and a new leptonic generation has taken its place for anomaly cancellation purposes. 
This introduction of a new leptonic generation  would have significant effects in 
the Higgs production rates and eventually will change the SM expectations. 
In the second model, the  second Higgs superfield of the MSSM is turned into a spectator. In both cases,
mass generation can be implemented by the use of $H_u$ and the constrained superfield $X$.  
It should be noted that in both cases the $\mu$ problem of the MSSM does not exist, in the first model by construction 
(as there is no $H_d$) and in the second case by the employment
of a discrete symmetry.

Thus, one can have a non-linear MSSM where there is only one field with the ``Higgs'' 
property (i.e., of getting a vacuum expectation value). 
The constrained superfield framework we used, especially the goldstino, 
which should be interpreted as the surviving longitudinal low energy component of the gravitino, 
gives an insight to the connection of the more fundamental supergravity theory with the low energy phenomenology. 
We stress again that, it is in this sense that the supersymmetric single-Higgs Yukawa couplings 
are fundamentally connected to the low energy limit of supergravity, rather than being completely 
unattached to this underlying theory.

We would like to make a final comment in the case of a half quark generation.
Electroweak symmetry breaking in the single Higgs non-linear MSSM should happen again radiatively. 
 Quantum corrections drive the initially positive soft mass of the Higgs field to negative values near 
the electroweak scale and thus triggers symmetry breaking. 
This happens due to the large Yukawa couplings of the Higgs field to matter, especially the heavy quarks. 
It will be the new generation heavy quarks that  will dominate radiative corrections 
and will make this effect quite stronger.


\section*{Acknowledgements}
The authors thank E. Dudas, G. Orfanidis and P. Tziveloglou for discussions and correspondence.
This research was implemented under the ``ARISTEIA" Action of the 
``Operational Programme Education and Lifelong Learning''
and is co-funded by the European 
Social Fund (ESF) and National Resources.

\newpage

\appendix

\section{Constrained Superfield}

The constrained superfield (goldstino) Lagrangian has the usual form \cite{Komargodski:2009rz}
\be
\label{XNL-appendix}
{\cal{L}}_{X}= \int d^4 \theta \bar{X} X + \Big\{ \int d^2 \theta f X  +h.c. \Big\}
\ee
with $f$ the hidden sector SUSY breaking scale. The superfield $X$ satisfies the constraint
\beq
\label{goldstino-constraint}
X^2=0
\eeq
This constraint gives a relation among the component fields
allowing to integrate out the sgoldstino in terms of the goldstino and the auxiliary field $F$, as
\beq
\label{goldstino-components}
\phi=\frac{GG}{2F}
\eeq
so that  the component Lagrangian is written as
\be
\label{XNL-appendix-components}
{\cal{L}}_{X}= i \partial_\mu \bar{G} \bar{\sigma} G + \bar{F} F 
+ \frac{\bar{G}^2}{2\bar{F}} \partial^2 \left( \frac{G^2}{2F}  \right)
+ \Big\{ f F +h.c. \Big\}.
\ee
The equations of motion for the auxiliary field $F$ (and $\bar{F}$) read
\be
\label{Feq}
\nonumber
F + f - \frac{\bar{G}^2}{2 \bar{F}^2} \partial^2 \left(\frac{G^2}{2 F}\right) &=&0,
\\
\bar{F} + f - \frac{G^2}{2 F^2} \partial^2 \left(\frac{\bar{G}^2}{2 \bar{F}}\right) &=&0
\ee
which are solved by
\be
\label{Fsol}
\nonumber
F&=& - f \left( 1 + \frac{\bar{G}}{4 f^4}  \partial^2  G^2 -\frac{3}{16f^8} G^2 \bar{G}^2  \partial^2 G^2 \partial^2 \bar{G}^2 \right),
\\
\bar{F}&=& - f \left( 1 + \frac{G}{4 f^4}  \partial^2  \bar{G}^2 -\frac{3}{16f^8} G^2 \bar{G}^2  \partial^2 G^2 \partial^2 \bar{G}^2 \right)
\ee
Inserting (\ref{Fsol}) back into (\ref{XNL-appendix-components}) the  on-shell Lagrangian 
\be
\label{XNL-appendix-components-on-shell}
{\cal{L}}_{X}= -f^2 + i \partial_\mu \bar{G} \bar{\sigma} G 
+ \frac{1}{4 f^2} \bar{G}^2 \partial^2 G^2 
-\frac{1}{16 f^6}G^2 \bar{G}^2 \partial^2 G^2\partial^2\bar{G}^2.
\ee
is recovered. Note that (\ref{XNL-appendix-components-on-shell})
 is equivalent to the well known Akulov-Volkov Lagrangian \cite{Volkov:1973ix}.

\section{Higher Dimensional Operators}

We present the higher dimensional operators that serve as the building block for the component form of the Lagrangians (\ref{Yd'-superspace}) and (\ref{td'-superspace}). The component Lagrangian for the Yukawa couplings is
\be
\label{HDO1-components-Appendix}
\nonumber
 {\cal{L}}_{Y} &= &- \frac{m_{soft}}{f\, \Lambda} \int d^2 \theta d^2  \bar{\theta} \bar{X} \bar{H} e^V Q \bar{d}
\\
\nonumber
&=& - \frac{m_{soft}}{f\, \Lambda}\, \left\{ - \bar{F} (   \bar{h}^+, \bar{h}^0 ) \left( \begin{array}{c} u_{L}^{\alpha} \\ d_{L}^{\alpha} \end{array} \right) \bar{d}_{L \alpha} 
+  \bar{F} (   \bar{h}^+, \bar{h}^0 )  \left( \begin{array}{c} \tilde{u}_{L} \\ \tilde{d}_{L} \end{array} \right)  F_{\bar{d}_L}
\right.
\\
\nonumber
&&+ \bar{F} (   \bar{h}^+, \bar{h}^0 )   \left( \begin{array}{c} F_{u_L} \\ F_{ d_L}  \end{array} \right)  \tilde{\bar{d}}_L 
+ i \partial_a \bar{G}_{\dot{\rho}} \bar{\sigma}^{a \dot{\rho} \alpha} (   \bar{h}^+, \bar{h}^0 ) \left( \begin{array}{c} u_{L\alpha} \\ d_{L\alpha} \end{array} \right) \tilde{\bar{d}}_L
\\
\nonumber
&&+ i \partial_a \bar{G}_{\dot{\rho}} \bar{\sigma}^{a \dot{\rho} \alpha} (   \bar{h}^+, \bar{h}^0 ) \left( \begin{array}{c} \tilde{u}_{L} \\ \tilde{d}_{L} \end{array} \right) \bar{d}_{L \alpha} 
+ \Box \bar{\phi} (   \bar{h}^+, \bar{h}^0 ) \left( \begin{array}{c} \tilde{u}_L \\ \tilde{d}_L \end{array} \right) \tilde{\bar{d}}_L
\\
\nonumber
&&+2 \partial_a \bar{\phi} [ \partial^a (   \bar{h}^+, \bar{h}^0 )  ]  \left( \begin{array}{c} \tilde{u}_L \\ \tilde{d}_L \end{array} \right)  \tilde{\bar{d}}_L 
+ i \sigma^{a}_{\alpha\dot{\alpha}} \partial_a \bar{\phi} ( \bar{\tilde{h}}^{+\dot{\alpha}}, \bar{\tilde{h}}^{0\dot{\alpha}}) \left( \begin{array}{c} u_{L}^{\alpha} \\ d_{L}^{\alpha} \end{array} \right) \tilde{\bar{d}}_L
\\
\nonumber
&&+ i \sigma^{a}_{\alpha\dot{\alpha}} \partial_a \bar{\phi} ( \bar{\tilde{h}}^{+\dot{\alpha}}, \bar{\tilde{h}}^{0\dot{\alpha}}) \left( \begin{array}{c} \tilde{u}_{L} \\ \tilde{d}_{L} \end{array} \right) \bar{d}_{L}^{\alpha} 
+ i \bar{G}^{\dot{\alpha}} \sigma^{a}_{\alpha\dot{\alpha}} [ \partial_a (   \bar{h}^+, \bar{h}^0 )  ] \left( \begin{array}{c} u_{L}^{\alpha} \\ d_{L}^{\alpha} \end{array} \right) \tilde{\bar{d}}_L
\\
\nonumber
&&- \bar{G}_{\dot{\alpha}} ( \bar{\tilde{h}}^{+\dot{\alpha}}, \bar{\tilde{h}}^{0\dot{\alpha}}) \left( \begin{array}{c} F_{u_L} \\ F_{ d_L}  \end{array} \right)  \tilde{\bar{d}}_L + \bar{G}_{\dot{\alpha}} ( \bar{\tilde{h}}^{+\dot{\alpha}}, \bar{\tilde{h}}^{0\dot{\alpha}}) \left( \begin{array}{c} u_{L}^{\alpha} \\ d_{L}^{\alpha} \end{array} \right) \bar{d}_{L\alpha}
\\
\nonumber
&&+ i \bar{G}^{\dot{\alpha}} \sigma^{a}_{\alpha\dot{\alpha}} [ \partial_a (   \bar{h}^+, \bar{h}^0 )  ] \left( \begin{array}{c} \tilde{u}_{L} \\ \tilde{d}_{L} \end{array} \right) \bar{d}_{L}^{\alpha} 
- \bar{G}_{\dot{\alpha}} ( \bar{\tilde{h}}^{+\dot{\alpha}}, \bar{\tilde{h}}^{0\dot{\alpha}}) \left( \begin{array}{c} \tilde{u}_{L} \\ \tilde{d}_{L} \end{array} \right) F_{\bar{d}_{L}}
\\
\nonumber
&&- i \partial_a \bar{\phi} (   \bar{h}^+, \bar{h}^0 ) {\cal{V}}^a \left( \begin{array}{c} \tilde{u}_{L} \\ \tilde{d}_{L} \end{array} \right) \tilde{\bar{d}}_L 
- \frac{i}{\sqrt{2}} \bar{G}_{\dot{\alpha}} (   \bar{h}^+, \bar{h}^0 ) \bar{\lambda}^{\dot{\alpha}} \left( \begin{array}{c} \tilde{u}_{L} \\ \tilde{d}_{L} \end{array} \right) \tilde{\bar{d}}_L
\\
\nonumber
&&+  \frac{1}{2} \bar{\sigma}^{a \dot{\alpha} \alpha} \bar{G}_{\dot{\alpha}} (   \bar{h}^+, \bar{h}^0 ) {\cal{V}}_a \left( \begin{array}{c} u_{L\alpha} \\ d_{L\alpha} \end{array} \right) \tilde{\bar{d}}_L 
+ \frac{1}{2} \bar{\sigma}^{a \dot{\alpha} \alpha} \bar{G}_{\dot{\alpha}} (   \bar{h}^+, \bar{h}^0 ) {\cal{V}}_a \left( \begin{array}{c} \tilde{u}_{L} \\ \tilde{d}_{L} \end{array} \right) \bar{d}_{L\alpha}
\\
\nonumber
&&+ \bar{\phi} [ \Box (   \bar{h}^+, \bar{h}^0 )  ] \left( \begin{array}{c} \tilde{u}_{L} \\ \tilde{d}_{L} \end{array} \right) \tilde{\bar{d}}_L 
+ i \bar{\phi} \bar{\sigma}^{a \dot{\alpha} \alpha} [ \partial_a ( \bar{\tilde{h}}^{+}_{\dot{\alpha}}, \bar{\tilde{h}}^{0}_{\dot{\alpha}})  ] \left( \begin{array}{c} u_{L\alpha} \\ d_{L\alpha} \end{array} \right) \tilde{\bar{d}}_L
\\
\nonumber
&&+ i \bar{\phi} \bar{\sigma}^{a \dot{\alpha} \alpha} [ \partial_a ( \bar{\tilde{h}}^{+}_{\dot{\alpha}}, \bar{\tilde{h}}^{0}_{\dot{\alpha}})  ] \left( \begin{array}{c} \tilde{u}_{L} \\ \tilde{d}_{L} \end{array} \right) \bar{d}_{L \alpha} 
+ \bar{\phi}  ( \bar{F}^+, \bar{F}^0)   \left( \begin{array}{c} F_{u_L} \\ F_{ d_L}  \end{array} \right) \tilde{\bar{d}}_L
\\
\nonumber
&&+ \bar{\phi} ( \bar{F}^+, \bar{F}^0) \left( \begin{array}{c} u_{L\alpha} \\ d_{L\alpha} \end{array} \right) \bar{d}_{L}^{\alpha} 
+  \bar{\phi} ( \bar{F}^+, \bar{F}^0) \left( \begin{array}{c} \tilde{u}_{L} \\ \tilde{d}_{L} \end{array} \right)  F_{\bar{d}_L}
\\
\nonumber
&&- i \bar{\phi} [ \partial_a (   \bar{h}^+, \bar{h}^0 )  ] {\cal{V}}^a \left( \begin{array}{c} \tilde{u}_{L} \\ \tilde{d}_{L} \end{array} \right) \tilde{\bar{d}}_L 
- \frac{i}{\sqrt{2}}  \bar{\phi} ( \bar{\tilde{h}}^{+}_{\dot{\alpha}}, \bar{\tilde{h}}^{0}_{\dot{\alpha}}) \bar{\lambda}^{\dot{\alpha}} \left( \begin{array}{c} \tilde{u}_{L} \\ \tilde{d}_{L} \end{array} \right) \tilde{\bar{d}}_L
\\
\nonumber
&&+ \frac{1}{4} \bar{\phi} \bar{\sigma}^{a \dot{\alpha} \alpha} ( \bar{\tilde{h}}^{+}_{\dot{\alpha}}, \bar{\tilde{h}}^{0}_{\dot{\alpha}}) {\cal{V}}_a \left( \begin{array}{c} u_{L\alpha} \\ d_{L\alpha} \end{array} \right) \tilde{\bar{d}}_L 
+ \frac{1}{4} \bar{\phi} \bar{\sigma}^{a \dot{\alpha} \alpha} ( \bar{\tilde{h}}^{+}_{\dot{\alpha}}, \bar{\tilde{h}}^{0}_{\dot{\alpha}}) {\cal{V}}_a \left( \begin{array}{c} \tilde{u}_{L} \\ \tilde{d}_{L} \end{array} \right) \bar{d}_{L \alpha}
\\
\nonumber
&&- \frac{i}{2} \bar{\phi} (   \bar{h}^+, \bar{h}^0 ) \partial_a {\cal{V}}^a \left( \begin{array}{c} \tilde{u}_{L} \\ \tilde{d}_{L} \end{array} \right) \tilde{\bar{d}}_{L} 
+ \frac{i}{\sqrt{2}} \bar{\phi} (   \bar{h}^+, \bar{h}^0 ) \lambda^\alpha \left( \begin{array}{c} u_{L\alpha} \\ d_{L\alpha} \end{array} \right) \tilde{\bar{d}}_{L}
\\
&&+ 
\left.\frac{i}{\sqrt{2}} \bar{\phi} (   \bar{h}^+, \bar{h}^0 ) \lambda^\alpha \left( \begin{array}{c} \tilde{u}_{L} \\ \tilde{d}_{L} \end{array} \right) \bar{d}_{L \alpha} - \frac{1}{4} \bar{\phi} (   \bar{h}^+, \bar{h}^0 ) [ {\cal{V}}^a {\cal{V}}_a 
-2 { \cal{D}} ] \left( \begin{array}{c} \tilde{u}_{L} \\ \tilde{d}_{L} \end{array} \right)  \tilde{\bar{d}}_{L} \  \right\}
\ee
where the gauge vector ${\cal{V}}^a$, the gaugino spinor ${\lambda}^{\alpha}$ and the auxiliary scalar ${ \cal{D}}$ of the gauge vector multiplet are Lie algebra valued. The component Lagrangian for the triple scalar couplings is
\be
\label{HDO2-components-Appendix}
\nonumber
 {\cal{L}}_{t}& =& - \frac{m_{soft}^2}{f^2\, \Lambda^2}\,  \int d^2 \theta d^2  \bar{\theta} \bar{X} X \bar{H} e^V Q \bar{d}
\\
\nonumber
&= &- \frac{1}{4 \sqrt{2} }\, \frac{m_{soft}^2}{f^2\, \Lambda^2}\,  G^{\alpha} \left\{ 
4i \bar{\phi} ( \bar{h}^+, \bar{h}^0 ) \lambda_{\alpha} \left( \begin{array}{c} \tilde{u}_{L} \\ 
\tilde{d}_{L} \end{array} \right)  \tilde{\bar{d}}_{L}\right. 
\\
\nonumber
&&-4 \sqrt{2} \bar{F} ( \bar{h}^+, \bar{h}^0 ) \left( \begin{array}{c} u_{L\alpha} \\ d_{L\alpha} \end{array} \right)  \tilde{\bar{d}}_{L}
-4 \sqrt{2} \bar{F} ( \bar{h}^+, \bar{h}^0 ) \left( \begin{array}{c} \tilde{u}_{L} \\ \tilde{d}_{L} \end{array} \right) \bar{d}_{L \alpha}
\\
\nonumber
&&+4 i \sqrt{2} {\epsilon}_{\alpha \beta} ( \partial_a \bar{G}_{\dot{\rho}}  ) \bar{\sigma}^{a\dot{\rho}\beta } ( \bar{h}^+, \bar{h}^0 ) \left( \begin{array}{c} \tilde{u}_{L} \\ \tilde{d}_{L} \end{array} \right)  \tilde{\bar{d}}_{L} 
- 4 i \sqrt{2} \sigma_{\alpha \dot{\alpha}}^{a} \partial_a \bar{\phi} ( \bar{\tilde{h}}^{+\dot{\alpha}}, \bar{\tilde{h}}^{0\dot{\alpha}} ) \left( \begin{array}{c} \tilde{u}_{L} \\ \tilde{d}_{L} \end{array} \right)  \tilde{\bar{d}}_{L} 
\\
\nonumber
&&- 4 i \sqrt{2} \bar{G}^{\dot{\alpha}} \sigma_{\alpha \dot{\alpha}}^{a} \partial_a ( \bar{h}^+, \bar{h}^0 ) \left( \begin{array}{c} \tilde{u}_{L} \\ \tilde{d}_{L} \end{array} \right)  \tilde{\bar{d}}_{L} 
- 4 \sqrt{2} \bar{G}^{\dot{\alpha}} ( \bar{\tilde{h}}^{+}_{\dot{\alpha}}, \bar{\tilde{h}}^{0}_{\dot{\alpha}} ) \left( \begin{array}{c} u_{L\alpha} \\ d_{L\alpha} \end{array} \right)  \tilde{\bar{d}}_{L} 
\\
\nonumber
&&- 4 \sqrt{2} \bar{G}^{\dot{\alpha}} ( \bar{\tilde{h}}^{+}_{\dot{\alpha}}, \bar{\tilde{h}}^{0}_{\dot{\alpha}} ) \left( \begin{array}{c} \tilde{u}_{L} \\ \tilde{d}_{L} \end{array} \right)  \bar{d}_{L \alpha} 
-2 \sqrt{2} \sigma_{\alpha \dot{\alpha}}^{a} \bar{G}^{\dot{\alpha}} ( \bar{h}^+, \bar{h}^0 ) {\cal{V}}_a \left( \begin{array}{c} \tilde{u}_{L} \\ \tilde{d}_{L} \end{array} \right)  \tilde{\bar{d}}_{L} 
\\
\nonumber
&&+ 4 i \sqrt{2} \bar{\sigma}^{a\dot{\rho}\beta } \bar{\phi} {\epsilon}_{\alpha \beta} \partial_a ( \bar{\tilde{h}}^{+}_{\dot{\rho}}, \bar{\tilde{h}}^{0}_{\dot{\rho}} ) \left( \begin{array}{c} \tilde{u}_{L} \\ \tilde{d}_{L} \end{array} \right)  \tilde{\bar{d}}_{L} 
- 4 \sqrt{2} \bar{\phi} ( \bar{F}^+ ,\bar{F}^0 ) \left( \begin{array}{c} u_{L\alpha} \\ d_{L\alpha} \end{array} \right)  \tilde{\bar{d}}_{L} 
\\
\nonumber
&&- 4 \sqrt{2} \bar{\phi} ( \bar{F}^+ ,\bar{F}^0 ) \left( \begin{array}{c} \tilde{u}_{L} \\ \tilde{d}_{L} \end{array} \right)  \bar{d}_{L \alpha} 
-2 \sqrt{2} \sigma_{\alpha \dot{\alpha}}^{a} \bar{\phi} ( \bar{\tilde{h}}^{+\dot{\alpha}}, \bar{\tilde{h}}^{0\dot{\alpha}} ){\cal{V}}_a \left( \begin{array}{c} \tilde{u}_{L} \\ \tilde{d}_{L} \end{array} \right)  \tilde{\bar{d}}_{L} \Big\}
\\
\nonumber
&&\left.+\frac{m_{soft}^2}{f^2\, \Lambda^2}\, F  \Big\{ - \bar{F} ( \bar{h}^+, \bar{h}^0 ) \left( \begin{array}{c} \tilde{u}_{L} \\ \tilde{d}_{L} \end{array} \right)  \tilde{\bar{d}}_{L} 
+ \bar{G}_{\dot{\alpha}} ( \bar{\tilde{h}}^{+}_{\dot{\alpha}}, \bar{\tilde{h}}^{0}_{\dot{\alpha}} )  \left( \begin{array}{c} \tilde{u}_{L} \\ \tilde{d}_{L} \end{array} \right) \tilde{\bar{d}}_{L}
- \bar{\phi} ( \bar{F}^+ ,\bar{F}^0 )  \left( \begin{array}{c} \tilde{u}_{L} \\ 
\tilde{d}_{L} \end{array} \right) \tilde{\bar{d}}_{L}   \right\}
\\
&&+ \frac{m_{soft}}{f\, \Lambda} \phi \ {\cal{L}}_{Y}.
\ee
In  (\ref{HDO1-components-Appendix}) and (\ref{HDO2-components-Appendix}) the sgoldstino has not yet been integrated out.
When this is done (by using $\phi=\frac{GG}{2F}$), a number of further goldstino couplings will appear.

\end{document}